# Exploring EEG-driven brain-heart coupling across sleep stages in individuals with sleep disorders

Jathushan Kaetheeswaran[1,2,3], Jenny Wei[1,4]

[1] Institute of Biomedical Engineering, University of Toronto, Toronto, ON, Canada
[2] KITE Research Institute, Toronto Rehabilitation Institute, University Health Network, Toronto, ON, Canada
[3] Krembil Brain Institute, University Health Network, Toronto, ON, Canada
[4] Centre for Digital Therapeutics, University Health Network, Toronto, ON, Canada

## ABSTRACT

The interactions between the brain and heart during sleep are responsible for regulating autonomic function. While brain-heart coupling has been studied in healthy populations, the relationships between neural and cardiac activity across sleep stages in the presence of sleep disorders are not clear. This study examines the influence of brain-driven cardiac activity across sleep stages for individuals with sleep disorders. Overnight recordings of C3 and C4 electroencephalogram (EEG) channels and electrocardiogram (ECG) signals from 146 individuals were preprocessed and analyzed in the frequency domain through a linear mixed-effect model. Our results show that parasympathetic activity is sensitive to changes in delta and beta powers during later stages of non-rapid eye movement (NREM) sleep, as both band powers exhibited strong negative effects on high-frequency heart rate variability (HF-HRV) power. These findings show that neural activity can drive vagal tone across sleep stages, suggesting that treatments on key EEG bands during NREM and REM stages may help restore regular cardiac behaviour.

## 1. BACKGROUND

Sleep disorders significantly elevate the risk of cardiovascular and neurological diseases. For instance, obstructive sleep apnea, a common sleep disorder among older adults, has been shown to double the risk of cardiovascular mortality and stroke [1], [2]. These disorders disrupt normal sleep architecture, which is comprised of the following stages: wakefulness, non-rapid eye movement (NREM), and REM sleep. Frequent disruptions negatively affect the regular autonomic regulations between neural and cardiac activity. This connection, known as brain-heart coupling, has demonstrated strong correlations with autonomic activity during sleep in healthy populations, specifically showing the strongest information transfer from the heart to the brain [3], [4]. While heart-to-brain information transfer dominates during healthy sleep, brain-to-heart modulations remain present and are sensitive to spectral band powers and cortical sites [4]. This brain-driven cardiac activity may exhibit a modified relationship in individuals with sleep disorders, however, this coupling has not been thoroughly studied across all sleep stages for this population. This study investigates if unit increases in neural power exhibit statistically significant effects across sleep stages on parasympathetic activity in the presence of sleep disorders that differ from healthy sleep behaviour. The cortical target for this study is the sensorimotor cortex, based on existing sleep research showing that it is significantly associated with sleep disruptions at various stages, making it a relevant target for further investigation [5], [6]. Electroencephalography (EEG) and electrocardiography (ECG) data from polysomnography (PSG) recordings will also be used to quantify brain-heart dynamics. This investigation can illuminate neural signatures of sleep disorders and inform treatments designed to target specific disruptions in sleep stages to regulate parasympathetic activity for improved restfulness.

**2. OBJECTIVE**

This study proposes that changes in EEG band power significantly affect HF power and that this relationship has abnormal behaviours in individuals with sleep disorders. We will investigate this through mixed-effects modelling that controls for subject variations and imbalanced data. We expect that brain-heart coupling will show significant differences compared to healthy sleep behaviours during restorative NREM sleep where slow-wave activity increases with vagal tone [3].

**3. METHODS**

*3.1. Dataset*

The Haaglanden Medisch Centrum sleep database was used to acquire 151 PSG recordings, which include four EEG channels (F4, C, O2/M1, and C3/M2), one ECG channel, and 30-second sleep stage annotations by sleep technicians [7], [8]. The dataset is comprised of patients referred to the sleep clinic in 2018 for sleep disorders sampled at 256 Hz. We used the C3 and C4 electrodes for analysis, as these represent the left and right primary motor cortex hemispheres, respectively. Subjects SN028, SN036, SN098, SN111, and SN115 were excluded due to corrupt ECG data. Sleep stages 0, 1, 2, 3 and 4 correspond to wake, NREM 1, NREM 2, NREM 3, and REM stages respectively.

*3.2. Heart Preprocessing*

Heart rate variability (HRV) is a standard measure of autonomic activity from ECG data. For HRV spectral analysis, high-frequency (HF) power (0.15-0.4Hz) is accepted to be a strong representation of parasympathetic activity in humans. Many sleep studies still use low-frequency (LF) power (0.04-0.15Hz) and LF: HF ratios despite extensive research disproving that LF power represents sympathetic activity [9-12]. To avoid conclusions based on false premises, LF powers were not analyzed. A median filter was applied to remove baseline wander noise (0.8-1.25 Hz), followed by moving window normalization using min-max normalization [13]. Heartbeat detection for each 30-second window was conducted using the Pan & Tompkins 1985 algorithm and the *SleepECG* library. The R-to-R interval (RRI) time series was computed per window and linearly interpolated at 4 Hz. A 4th-order Butterworth bandpass filter with a 0.04-0.4Hz band was applied to isolate relevant frequencies. A maximal discrete wavelet packet transform (MODWPT) was applied with a Daubechies 2 (*db2*) wavelet to decompose each 30-second window into frequency bands. Discrete wavelet transforms (DWT) are commonly utilized for power analysis of RRI signals, and MODWPTs benefit from better time resolution due to zero downsampling [14-17]. Ultra-short-term HF spectral analysis has previously been shown to be reliable under resting conditions for estimating cardiac features [18], [19]. Each terminal node of this decomposition represents the power of its frequency band. HF power was computed by calculating the fractional sums of relevant nodes and normalized to eliminate intersubject power variability [20].

*3.3. Brain Preprocessing*

The EEG band powers for C3 and C4 were computed in 30-second intervals using the *YASA* sleep analysis Python library. We applied a median Welch method to calculate band powers without amplifying the effect of potential outliers in each window [21]. Each window produced four spectral band values: delta (1-4Hz), theta (4-8Hz), alpha (8-12Hz), beta (12-30Hz), and gamma (30-80Hz).

*3.4. Statistical Analysis*

A linear mixed-effect model assessed the relationship between EEG and sleep stages on parasympathetic activity while accounting for subject variability. Repeated measures ANOVA was not chosen due to unbalanced repeated measures from varying sleep durations, as shown in Figure 1. The data was organized into long format and analyzed using JMP Pro. The linear mixed model assumed normally distributed residuals and random-effect coefficients [22].

$$HF\_Power_{norm} \sim EEG + Sleep\_Stage + (EEG \times Sleep\_Stage) + (1\,|\,Subject) + (1\,|\,Subject\!:\!Sleep\_Stage) \qquad (1)$$

We fit the model using EEG band powers, sleep stages, and their interactions as fixed effects. Random effects were set for subjects and sleep stages were nested within subjects. These random intercepts account for inter- and intra-subject variability across sleep stages. The final model equation can be seen in (1). A Yeo-Johnson transform normalized HF power before modelling, and conditional residuals were checked for violations of assumptions.

*3.5. Clustering to Investigate EEG-HRV Coupling Subtypes*

To explore subtypes in EEG-HRV coupling, we performed unsupervised hierarchical clustering on the preprocessed 30-second epochs. The number of clusters was chosen based on the dendrogram. Principal component analysis (PCA) was applied for two-dimensional visualization of cluster separability. For each cluster, we computed average feature values and conducted Tukey's Honest Significant Difference (HSD) test, comparing each cluster to every other cluster for each feature. We also examined subject-level cluster distributions to visualize the proportions of time subjects spent in each cluster to assess coupling variability.

## 4. RESULTS

*4.1. Descriptive Statistics*

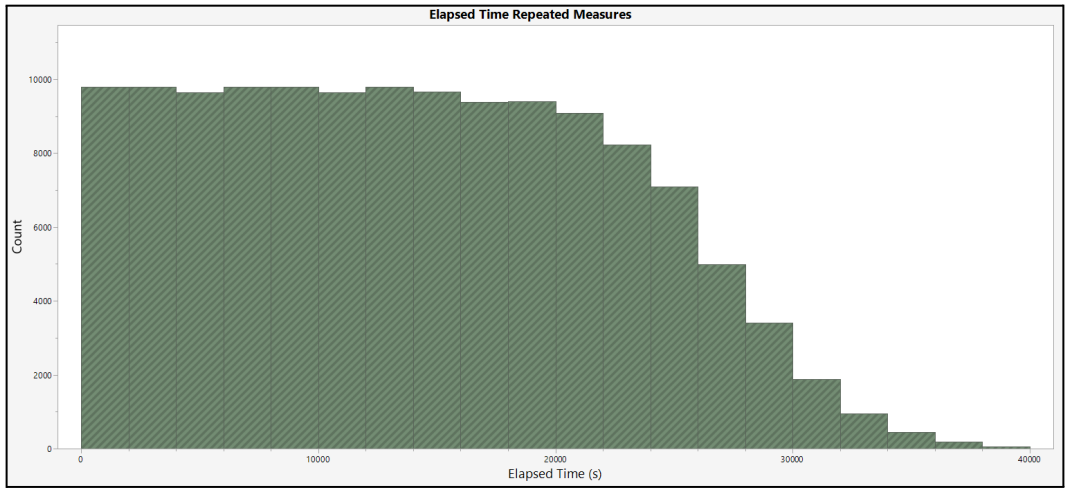

**Figure 1.** Histograms for elapsed time measures to illustrate uneven sleep durations. The unbalanced nature of these repeated measurements implies that traditional ANOVA analysis does not work for this dataset.

(a)

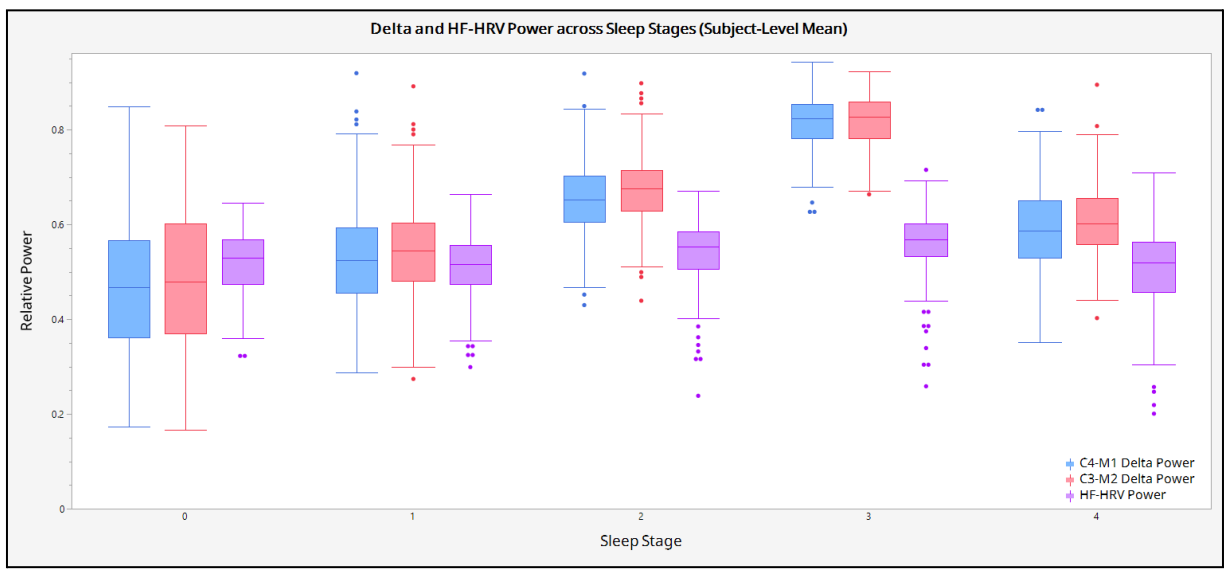

(b)

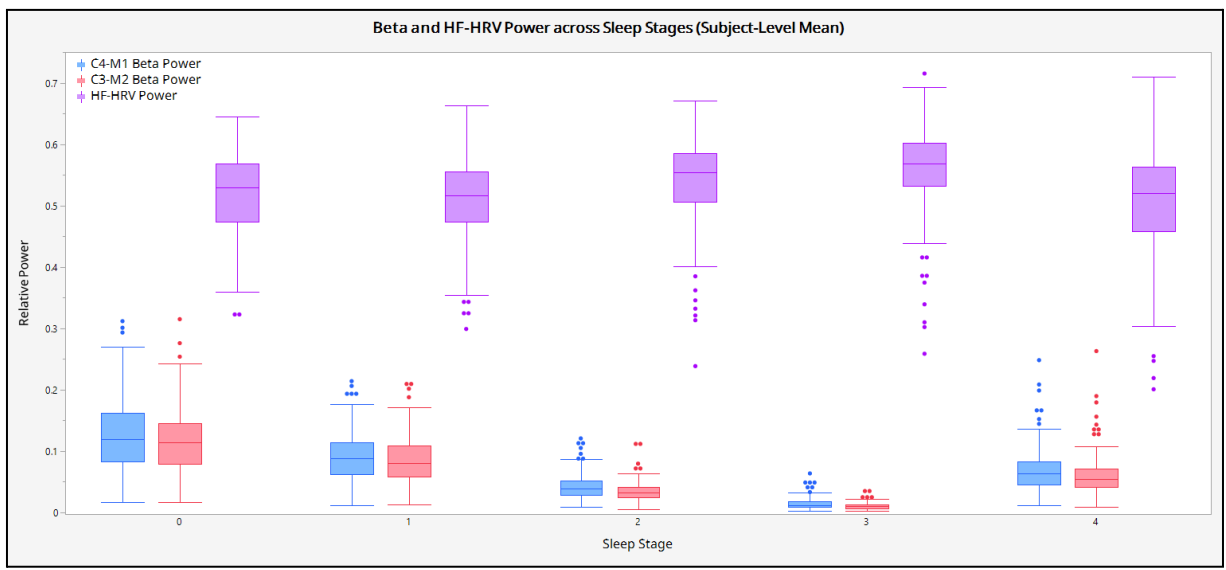

**Figure 2.** Boxplots of subject-level mean for a) delta power and b) beta power across sleep stages when compared to normalized HF-HRV power.

*4.2 Linear Mixed-Effect Model - Residual and Random-Effect Coefficients*

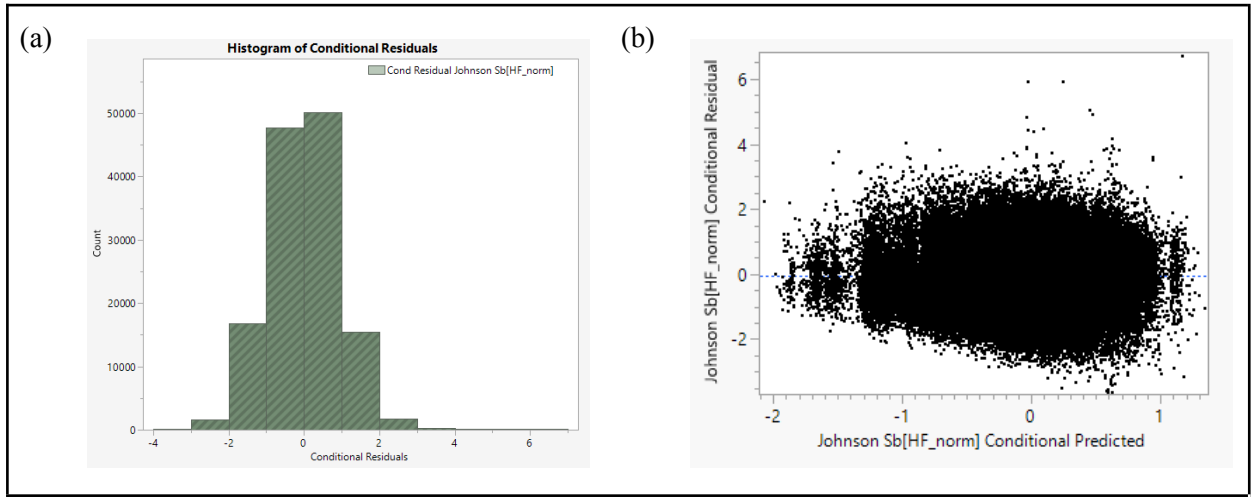

**Figure 3.** Normality plots for residuals through a) histogram analysis and b) scatterplot analysis. The results do not indicate severe violations of assumptions for linear mixed-effect modelling.

**Table 1.** Statistically significant ($p < 0.05$) fixed effect parameter estimates using indicator coding. The *Estimate* value represents the direction and magnitude of change in our response variable, transformed HF power, per unit increase in EEG band power per sleep stage.

| Sleep Stage | Electrode | EEG Band | Estimate | Std Error | t Ratio | Prob > \|t\| |
|---|---|---|---|---|---|---|
| NREM 2 (N2) | C4 | Delta | -1.810349 | 0.5657814 | -3.20 | 0.0014 |
| | | Theta | -2.488999 | 0.6257574 | -3.98 | <.0001 |
| | | Alpha | -1.700986 | 0.6824431 | -2.49 | 0.0127 |
| | | Beta | -2.958661 | 0.9309039 | -3.18 | 0.0015 |
| | | Gamma | -1.934917 | 0.9853582 | -1.96 | 0.0496 |
| | C3 | Delta | -2.841344 | 0.5340294 | -5.32 | <.0001 |
| | | Theta | -2.849056 | 0.5889885 | -4.84 | <.0001 |
| | | Alpha | -3.108054 | 0.6389435 | -4.86 | <.0001 |
| | | Beta | -7.026271 | 0.8902143 | -7.89 | <.0001 |
| | C4 | Delta | -2.492933 | 1.2190178 | -2.05 | 0.0409 |
| | | Theta | -2.689199 | 1.3048509 | -2.06 | 0.0393 |
| NREM 3 (N3) | | Beta | -7.975946 | 2.3559647 | -3.39 | 0.0007 |
| | C3 | Delta | -3.677407 | 1.1589181 | -3.17 | 0.0015 |
| | | Theta | -4.099152 | 1.2410277 | -3.30 | 0.001 |
| | | Alpha | -3.476499 | 1.4178707 | -2.45 | 0.0142 |
| | | Beta | -13.91606 | 2.403097 | -5.79 | <.0001 |
| | | Gamma | 6.339592 | 3.2319526 | 1.96 | 0.0498 |
| REM | C4 | Delta | -2.516995 | 0.8755232 | -2.87 | 0.004 |
| | | Theta | -2.375782 | 0.9333908 | -2.55 | 0.0109 |
| | | Alpha | -3.084737 | 1.0616216 | -2.91 | 0.0037 |
| | | Beta | -4.197251 | 1.3404465 | -3.13 | 0.0017 |

*4.3 Clustering Analysis of NREM2*

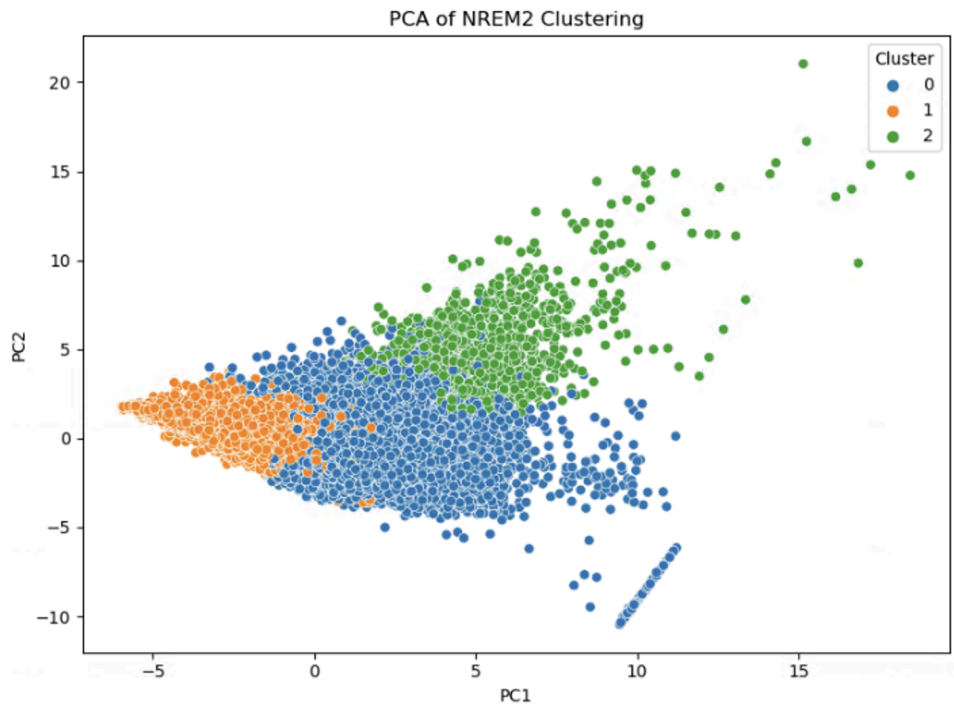

**Figure 4.** PCA Visualization of Hierarchical Clustering of NREM2.

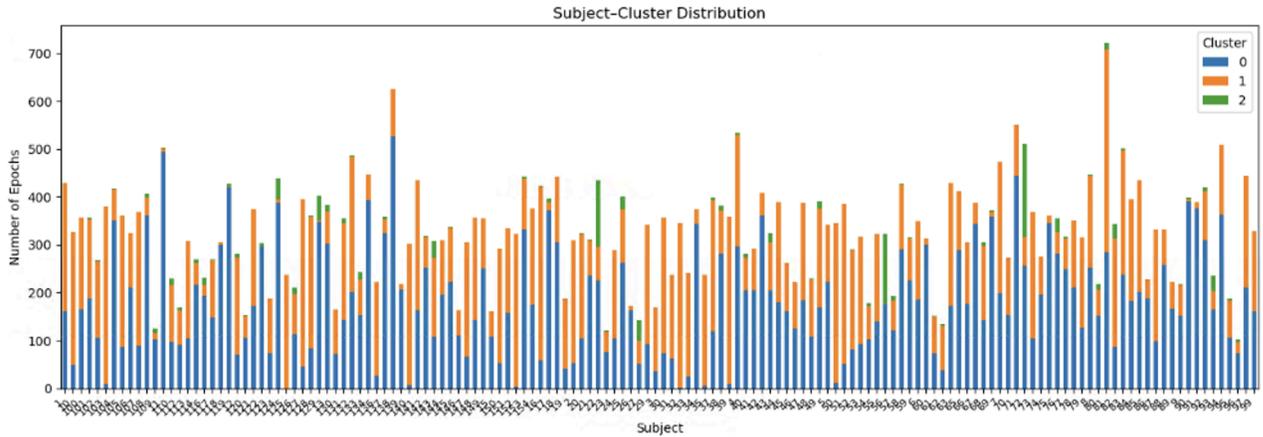

**Figure 5.** Subject-Cluster Distribution for NREM2: The stacked bars represent the proportion of time each subject spends in each cluster, providing insight into the variability of brain-heart coupling profiles within the cohort.

**Table 2.** Feature Means of NREM2 Clusters.

| Electrode | EEG Band | Cluster 0 | Cluster 1 | Cluster 2 |
|---|---|---|---|---|
| C3-M2 | Alpha | 0.09875076 | 0.05459761 | 0.099492305 |
| | Beta | 0.04108835 | 0.02129415 | 0.141520582 |
| | Delta | 0.60762219 | 0.75124535 | 0.489971261 |
| | Gamma | 0.00743196 | 0.00398362 | 0.043039187 |
| | Theta | 0.1925898 | 0.13502187 | 0.15077568 |
| C4-M1 | Alpha | 0.10922886 | 0.05732277 | 0.106573766 |
| | Beta | 0.05102287 | 0.02537404 | 0.141295099 |
| | Delta | 0.57727487 | 0.74509168 | 0.477065254 |
| | Gamma | 0.00887514 | 0.00414857 | 0.042248762 |
| | Theta | 0.19513648 | 0.13242734 | 0.152790784 |
| HF | | 295450.348 | 733367.595 | 177644.9092 |

**Table 3**. Top five statistically significant features of NREM2 clusters based on Tukey's HSD post-hoc test.

| Cluster | Top Discriminative Features |
|---|---|
| 0 | ↓ HF, ↑ C4-M1 Theta, ↑ C3-M2 Theta, ↓ C3-M2 Beta, ↓ C4-M1 Delta |
| 1 | ↑ HF, ↑ C4-M1 Delta, ↑ C3-M2 Delta, ↓ C4-M1 Beta, ↓ C3-M2 Beta |

| 2 | ↓ HF, ↓ C3-M2 Delta, ↓ C4-M1 Delta, ↑ C3-M2 Beta, ↑ C4-M1 Beta |
|---|---|

*4.4 Clustering Analysis of REM*

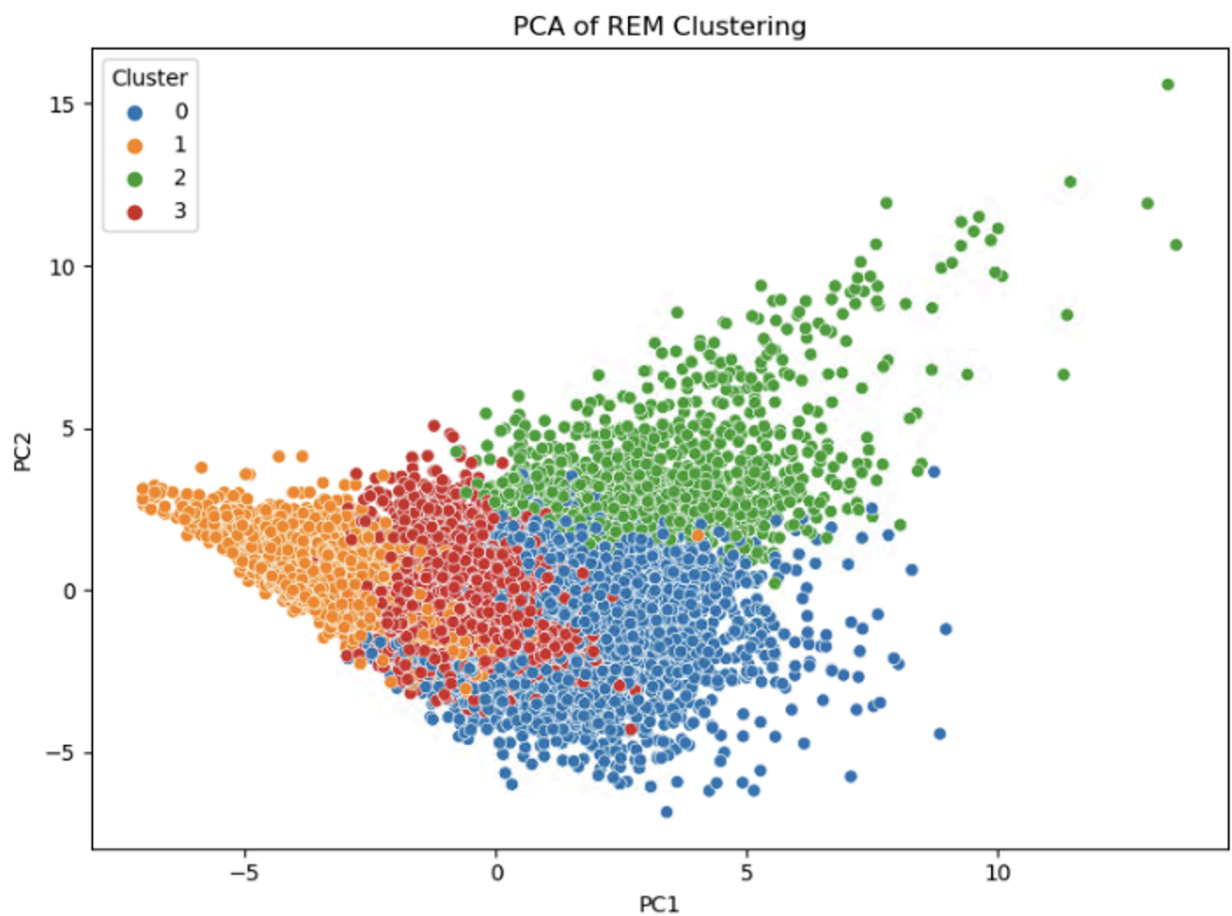

**Figure 6.** PCA Visualization of hierarchical clustering of REM sleep.

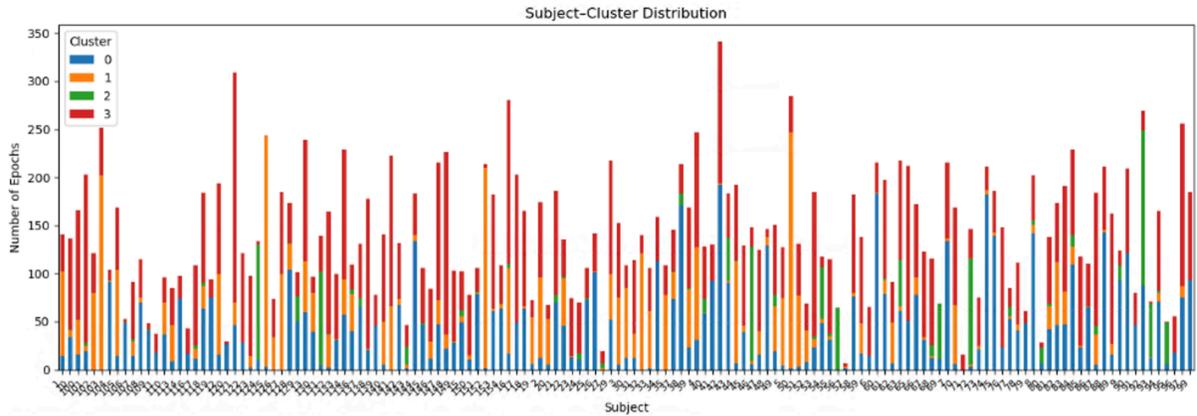

**Figure 7.** Subject-cluster distribution for REM sleep.

**Table 4.** Feature means of REM clusters.

| Electrode | EEG Band | Cluster 0 | Cluster 1 | Cluster 2 | Cluster 3 |
|-----------|----------|-----------|-----------|-----------|-----------|
| C3-M2 | Alpha | 0.10974257 | 0.05119944 | 0.08096864 | 0.07867454 |
|  | Beta | 0.07212495 | 0.0316389 | 0.14466525 | 0.0485583 |
|  | Delta | 0.52084994 | 0.72385014 | 0.52388075 | 0.62352896 |
|  | Gamma | 0.00986663 | 0.00524652 | 0.03309172 | 0.00785729 |
|  | Theta | 0.23716049 | 0.16497389 | 0.1649142 | 0.20586241 |
| C4-M1 | Alpha | 0.11764168 | 0.05226081 | 0.09269632 | 0.0824341 |
|  | Beta | 0.08099008 | 0.03470357 | 0.15487768 | 0.05594164 |
|  | Delta | 0.50313271 | 0.73645397 | 0.49460474 | 0.61689791 |
|  | Gamma | 0.00995156 | 0.00478042 | 0.0267843 | 0.00796148 |
|  | Theta | 0.22979879 | 0.14538646 | 0.16507871 | 0.19532636 |
| HF |  | 400296.624 | 1043492.64 | 116925.878 | 277285.76 |

**Table 5.** Top five statistically significant features of REM Clusters based on Tukey's HSD post-hoc test.

| Cluster | Top Discriminative Features |
|---------|-----------------------------|
| 0 | ↓ HF, ↓ C4-M1 Delta, ↑ C4-M1 Theta, ↑ C3-M2 Theta, ↑ C4-M1 Alpha |
| 1 | ↑ HF, ↑ C4-M1 Delta, ↑ C3-M2 Delta, ↓ C4-M1 Beta, ↓ C3-M2 Beta |
| 2 | ↓ HF, ↓ C4-M1 Delta, ↑ C4-M1 Beta, ↑ C3-M2 Beta, ↑ C3-M2 Gamma |
| 3 | ↓ HF, ↑ C4-M1 Delta, ↓ C3-M2 Beta, ↓ C4-M1 Beta, ↑ C3-M2 Delta |

## 5. DISCUSSION & IMPLICATIONS

The mixed-effects model results show no severe assumption violations (Figure 3), indicating reliable estimates. Although the residuals slightly deviate from normality, linear mixed-effects models are resilient to deviations [22]. The significant fixed effect estimates (Table 1) are interpreted as follows: the estimate corresponds to the change in HF power per unit increase in the corresponding EEG band power and sleep stage. The t-test is used to determine the effect size while accounting for standard error. It is important to note that the NREM 1 sleep stage did not yield any significant changes from wake stage coupling. This is generally expected given that it is an unstable stage that is difficult to distinguish from the wake stage brain behaviour for even expert sleep technicians [23]. The key interpretations from the mixed model estimates are discussed below.

### 5.1. Cortical delta power suppresses parasympathetic activity during NREM sleep

Greater parasympathetic activity is linked to the NREM sleep stage, as shown in Figure 2a by the increase in HF power relative to other sleep stages [24]. The NREM stage is also characterized by significant increases in delta power, evident in the powers of both the left and right motor hemispheres shown in Figure 2a. In healthy individuals, HF power positively correlates with delta power, reflecting the dominance of slow-wave activity and vagal tone during deep sleep [25]. However, the results from this population indicate a negative coupling between cortical delta activity and parasympathetic activity. The effect was strongest in the left hemisphere during N2 (Estimate = -2.84 ± 0.53, t = -5.32, $p < .0001$), suggesting that higher cortical delta power significantly suppresses vagal tone. Prior research has demonstrated that HF-Delta power coherence during NREM becomes increasingly disrupted as the severity of sleep apnea rises [26]. Considering the presence of sleep disorders in this study, these findings may reflect a gradual shift in the relationship between slow-wave sleep and parasympathetic activity, evolving from positive coupling to negative coupling, and potentially to complete disruption as disorder severity increases.

### 5.2. Cortical beta power is strongly coupled with vagal tone during NREM sleep

While the group-level trends from Figure 2b show a general decrease in beta power from wake to NREM sleep, the results from our model show that any increases in individual beta power will significantly reduce parasympathetic activity. Although both hemispheres showed significant negative effects, the left hemisphere (C3) exhibited larger effects in both N2 (Estimate = -7.03 ± 0.89, t = -7.89, $p < .0001$) and N3 (Estimate = -13.92 ± 2.40, t = -5.79, $p < .0001$) compared to C4. These findings agree with sleep literature showing that beta power is a marker of cortical hyperarousal and sleep fragmentation, both of which decrease vagal tone and disrupt restorative sleep [27], [28]. However, studies with healthy populations have shown that HF-Beta power coupling is weak during NREM sleep [29]. The consistently strong negative effect of cortical beta power on vagal tone during NREM stages may indicate that this relationship is a physiological marker for poor sleep quality. This coupling may present a new treatment target to improve restorative sleep by suppressing beta activity during NREM stages and reducing cortical hyperarousal.

### 5.3. Parasympathetic activity is sensitive to C4 band powers during REM sleep

Autonomic control is significantly more variable during REM sleep than in the other sleep stages in healthy individuals. The decrease in vagal tone and increase in brain activity to facilitate dreaming suggest greater sympathetic dominance in this state [30]. The findings from our cohort align with this behaviour, as cortical powers from the right hemisphere (C4) all have a strong negative effect on HF power during REM sleep. Interestingly, only the right hemisphere's EEG activity had significant effects on vagal tone during this stage. This aligns with previous research that showed right hemisphere dominance in individuals with insomnia during REM sleep, suggesting that right-sided cortical activity drives brain-heart coupling for this stage [31]. Neuromodulatory research may benefit from targeting right-sided sensorimotor regions during REM sleep to examine the impact on REM stage sleep quality.

*5.4. Clustering Analysis*

The clustering analysis revealed distinct brain-heart coupling subtypes during NREM2 and REM sleep. In NREM2, Cluster 0 showed moderate theta activity but lower HF power, suggesting a transitional phase with lower parasympathetic activity. Cluster 1, characterized by low delta power across multiple electrodes, is indicative of a disrupted or shallow NREM2 phase. This may correspond to impaired autonomic regulation, reflecting potential sleep disturbances. Cluster 2 exhibited high delta power with strong parasympathetic activation, suggesting healthy NREM2 sleep. Cluster 3 exhibited elevated delta with increased beta and reduced HF power, suggesting impaired autonomic regulation and disrupted coupling.

In REM sleep, Cluster 0 displayed elevated HF power and delta with high theta, suggesting a balanced parasympathetic state with effective brain-heart coupling. Cluster 1, with increased delta and decreased HF, was associated with a restorative state. Cluster 2 presented high beta and gamma activity with decreased HF, indicating an arousal state and disrupted parasympathetic function. Finally, Cluster 3 showed high delta and reduced beta with low HF, indicating impaired brain-heart coupling and reduced autonomic regulation.

The clustering analysis and mixed effects modelling are consistent in identifying disrupted brain-heart coupling under conditions of hyperarousal (elevated beta power), with Cluster 1 in NREM2 and Cluster 3 in both NREM and REM showing disrupted autonomic regulation or parasympathetic function. The cluster analysis findings highlight the variability in autonomic regulation across both stages. The heterogeneity observed within subjects suggests that individuals with sleep disorders may transition between different coupling states, or remain within specific coupling profiles. The identification of specific coupling states in NREM2 and REM provides valuable phenotypes that can guide therapeutic strategies for improving sleep quality and regulating brain-heart coupling.

## 6. LIMITATIONS

This work has several limitations. It was not possible to perform analyses on subject demographics and comorbidities due to data anonymization of the provided dataset. Our analysis also calculated HF-HRV using 30-second windows to align with sleep-staging labels. Although previous literature has supported ultra-short-term HF-HRV spectral analysis as effective in static environments, the power estimates may still be susceptible to noise and fluctuations [18], [19]. Future work should examine whether the findings from this study are reproducible with more stable window sizes, such as five-minute segments. Additionally, the estimates from the linear mixed-effect model do not indicate the directionality of the coupling changes; rather, they show the variations of this relationship across sleep stages driven by neural activity changes. Transfer entropy analysis could be considered in future studies to investigate time-based brain-to-heart coupling and uncover the directionality of information transfer. Furthermore, this analysis is limited to a linear model, so future research may explore how nonlinear modelling fits this population.

## 6. CONCLUSION

This study investigated how brain-heart coupling varies across sleep stages in individuals with sleep disorders. Through applying linear mixed-effect modelling to EEG and ECG recordings, significant associations between EEG band power and HF power were found. Notably, delta and beta band power during NREM sleep were strongly negatively coupled with parasympathetic activity. Additionally, during REM sleep, only right hemisphere EEG activity significantly influenced HF power, suggesting lateralized brain-heart interactions in this stage. These findings differentiate from the patterns observed in healthy populations and highlight altered brain-heart coupling as an important physiological marker of sleep disorder. The clustering analysis identified distinct subtypes of brain-heart coupling across NREM and REM sleep stages. These clusters reflect variations in parasympathetic regulation and autonomic regulation, with important implications for understanding sleep disorders. Future work should explore additional nonlinear modelling to investigate targeted interventions for restoring healthy autonomic regulation through the modulation of specific EEG bands during sleep.

## 7. CONTRIBUTIONS

**Jathushan Kaetheeswaran**
Writing: Abstract, Background, Objective, Methods (3.1-3.4), Results (4.1-4.2), Discussion & Implications (5.1-5.3), Limitations, References
Technical work: Data extraction coding, Data preprocessing coding, Statistical experimentation, Statistical Analysis, Mixed-effects model interpretation and formatting

**Jenny Wei**
Writing: Methods (3.5), Results (4.3-4.4), Discussion & Implications (5.4), Conclusion
Technical work: Clustering analysis, Clustering results interpretation and formatting

# REFERENCES

[1]    R. Ravichandran, L. Gupta, M. Singh, A. Nag, J. Thomas, and B. K. Panjiyar, "The Interplay Between Sleep Disorders and Cardiovascular Diseases: A Systematic Review - PMC," Cureus, vol. 15, no. 9, Sept. 2023, doi: 10.7759/cureus.45898.

[2]    R. F. Gottesman, "Impact of Sleep Disorders and Disturbed Sleep on Brain Health: A Scientific Statement From the American Heart Association," Stroke, vol. 55, no.3, Mar. 2024, doi: 10.1161/STR.0000000000000453.

[3]    L. Faes, D. Marinazzo, F. Jurysta, and G. Nollo, "Linear and non-linear brain–heart and brain–brain interactions during sleep," Physiol. Meas., vol. 36, no. 4, pp. 683–698, Apr. 2015, doi: 10.1088/0967-3334/36/4/683.

[4]    J. Yang, Y. Pan, and Y. Luo, "Investigation of brain-heart network during sleep," in 2020 42nd Annual International Conference of the IEEE Engineering in Medicine & Biology Society (EMBC), Montreal, QC, Canada: IEEE, Jul. 2020, pp. 3343–3346. doi: 10.1109/EMBC44109.2020.9175305.

[5]    K. Bahr-Hamm, N. Koirala, M. Hanif, H. Gouveris, and M. Muthuraman, "Sensorimotor Cortical Activity during Respiratory Arousals in Obstructive Sleep Apnea," IJMS, vol. 24, no. 1, p. 47, Dec. 2022, doi: 10.3390/ijms24010047.

[6]    H. Gouveris et al., "Corticoperipheral neuromuscular disconnection in obstructive sleep apnoea," Brain Communications, vol. 2, no. 1, p. fcaa056, Jan. 2020, doi: 10.1093/braincomms/fcaa056.

[7]    Alvarez-Estevez, D., & Rijsman, R. (2022). Haaglanden Medisch Centrum sleep staging database (version 1.1). PhysioNet. https://doi.org/10.13026/t79q-fr32.

[8]    Alvarez-Estevez D, Rijsman RM (2021) Inter-database validation of a deep learning approach for automatic sleep scoring. PLoS ONE 16(8): e0256111. https://doi.org/10.1371/journal.pone.0256111

[9]    S. Michael, K. S. Graham, and G. M. Davis, "Cardiac Autonomic Responses during Exercise and Post-exercise Recovery Using Heart Rate Variability and Systolic Time Intervals—A Review," Front. Physiol., vol. 8, p. 301, May 2017, doi: 10.3389/fphys.2017.00301.

[10]    G. E. Billman, "The LF/HF ratio does not accurately measure cardiac sympatho-vagal balance," Front. Physio., vol. 4, 2013, doi: 10.3389/fphys.2013.00026.

[11]    B. L. Thomas, N. Claassen, P. Becker, and M. Viljoen, "Validity of Commonly Used Heart Rate Variability Markers of Autonomic Nervous System Function," Neuropsychobiology, vol. 78, no. 1, pp. 14–26, 2019, doi: 10.1159/000495519.

[12]    D. S. Goldstein, O. Bentho, M.-Y. Park, and Y. Sharabi, "Low-frequency power of heart rate variability is not a measure of cardiac sympathetic tone but may be a measure of modulation of cardiac

autonomic outflows by baroreflexes: Low-frequency power of heart rate variability," Experimental Physiology, vol. 96, no. 12, pp. 1255–1261, Dec. 2011, doi: 10.1113/expphysiol.2010.056259.

[13]     A. Paul, S. Panja, N. Das, and M. Mitra, "Median Filter Based Noise Reduction and QRS Detection in ECG Signal," in *Proceedings of International Conference on Industrial Instrumentation and Control*, vol. 815, S. Bhaumik, S. Chattopadhyay, T. Chattopadhyay, and S. Bhattacharya, Eds., in Lecture Notes in Electrical Engineering, vol. 815. , Singapore: Springer Nature Singapore, 2022, pp. 67–76. doi: 10.1007/978-981-16-7011-4_7.

[14]     P. Boudreau, W.-H. Yeh, G. A. Dumont, and D. B. Boivin, "Circadian Variation of Heart Rate Variability Across Sleep Stages," Sleep, vol. 36, no. 12, pp. 1919–1928, Dec. 2013, doi: 10.5665/sleep.3230.

[15]     C. A. García, A. Otero, X. Vila, and D. G. Márquez, "A new algorithm for wavelet-based heart rate variability analysis," 2014, doi: 10.48550/ARXIV.1411.5179.

[16]     U. Pale, F. Thurk, and E. Kaniusas, "Heart rate variability analysis using different wavelet transformations," in *2016 39th International Convention on Information and Communication Technology, Electronics and Microelectronics (MIPRO)*, Opatija: IEEE, May 2016, pp. 1649–1654. doi: 10.1109/MIPRO.2016.7522403.

[17]     A. Tzabazis, A. Eisenried, D. C. Yeomans, and M. I. Hyatt, "Wavelet analysis of heart rate variability: Impact of wavelet selection," *Biomedical Signal Processing and Control*, vol. 40, pp. 220–225, Feb. 2018, doi: 10.1016/j.bspc.2017.09.027.

[18]     J. W. Kim, H. S. Seok, and H. Shin, "Is Ultra-Short-Term Heart Rate Variability Valid in Non-static Conditions?," Front. Physiol., vol. 12, p. 596060, Mar. 2021, doi: 10.3389/fphys.2021.596060.

[19] D. Wehler et al., "Reliability of heart-rate-variability features derived from ultra-short ECG recordings and their validity in the assessment of cardiac autonomic neuropathy," Biomedical Signal Processing and Control, vol. 68, p. 102651, Jul. 2021, doi: 10.1016/j.bspc.2021.102651.

[20]     F. Shaffer and J. P. Ginsberg, "An Overview of Heart Rate Variability Metrics and Norms," Front. Public Health, vol. 5, p. 258, Sep. 2017, doi: 10.3389/fpubh.2017.00258.

[21]     L. Izhikevich, R. Gao, E. Peterson, and B. Voytek, "Measuring the average power of neural oscillations," Oct. 13, 2018, Neuroscience. doi: 10.1101/441626.

[22]     H. Schielzeth et al., "Robustness of linear mixed‑effects models to violations of distributional assumptions," Methods Ecol Evol, vol. 11, no. 9, pp. 1141–1152, Sep. 2020, doi: 10.1111/2041-210X.13434.

[23]     I. Lambert and L. Peter-Derex, "Spotlight on Sleep Stage Classification Based on EEG," Nature and Science of Sleep, vol. 15, pp. 479–490, Jun. 2023, doi: https://doi.org/10.2147/NSS.S401270.

[24]    M. Ako et al., "Correlation between electroencephalography and heart rate variability during sleep," Psychiatry and Clinical Neurosciences, vol. 57, no. 1, pp. 59–65, Feb. 2003, doi: https://doi.org/10.1046/j.1440-1819.2003.01080.x.

[25]    H. Abdullah, G. Holland, I. Cosic, and D. Cvetkovic, "Correlation of sleep EEG frequency bands and heart rate variability," PubMed, Sep. 2009, doi: https://doi.org/10.1109/iembs.2009.5334607.

[26]    Fabrice Jurysta et al., "The link between cardiac autonomic activity and sleep delta power is altered in men with sleep apnea-hypopnea syndrome," American Journal of Physiology-regulatory Integrative and Comparative Physiology, vol. 291, no. 4, pp. R1165–R1171, Oct. 2006, doi: https://doi.org/10.1152/ajpregu.00787.2005.

[27]    Y. Shi, R. Ren, F. Lei, Y. Zhang, M. V. Vitiello, and X. Tang, "Elevated beta activity in the nighttime sleep and multiple sleep latency electroencephalograms of chronic insomnia patients," Frontiers in Neuroscience, vol. 16, p. 1045934, 2022, doi: https://doi.org/10.3389/fnins.2022.1045934.

[28]    J. Fernandez-Mendoza et al., "Insomnia is Associated with Cortical Hyperarousal as Early as Adolescence," Sleep, vol. 39, no. 5, pp. 1029–1036, May 2016, doi: https://doi.org/10.5665/sleep.5746.

[29]    T. B. J. Kuo, C.-Y. Chen, Y.-C. Hsu, and C. C. H. Yang, "EEG beta power and heart rate variability describe the association between cortical and autonomic arousals across sleep," Autonomic Neuroscience, vol. 194, pp. 32–37, Jan. 2016, doi: 10.1016/j.autneu.2015.12.001.

[30]    I. M. Voronin and E. V. Biryukova, "Heart rate variability in healthy humans during night sleep," Human Physiology, vol. 32, no. 3, pp. 258–263, May 2006, doi: https://doi.org/10.1134/s0362119706030029.

[31]    G. V. Kovrov, S. I. Posokhov, and K. N. Strygin, "Interhemispheric EEG asymmetry in patients with insomnia during nocturnal sleep," Bulletin of Experimental Biology and Medicine, vol. 141, no. 2, pp. 197–199, Feb. 2006, doi: https://doi.org/10.1007/s10517-006-0126-z.